\documentclass[useAMS,graphics,usenatbib]{mn2e}
\usepackage{amssymb,color}
\usepackage{epsfig}

\newcommand{\galform}{\textsc{GALFORM}}

\newcommand{\dd}{{\rm d}}             
\newcommand{\mr}[1]{{\rm #1}}     

\newcommand{\nh}{n_\mr{H}}

\newcommand{\nhii}{n_\mr{HII}}

\title[Modelling recombinations during reionization]{Modelling recombinations during cosmological reionization}
\author[M. Rai\v{c}evi\'{c} and T. Theuns]{Milan Rai\v{c}evi\'{c}$^{1,2}$\thanks{E-mail:
milan.raicevic@durham.ac.uk} and Tom Theuns$^{1,3}$\\
$^1$Institute for Computational Cosmology, Durham University, Science Laboratories, Durham DH1 3LE,
UK\\
$^2$Leiden Observatory, Leiden University, P.O. Box 9513, 2300RA Leiden, The Netherlands\\
$^3$ Universiteit Antwerpen, Campus Groenenborger, Groenenborgerlaan
171, B-2020 Antwerpen, Belgium\\ 
}

\begin{document}

\date{}

\pagerange{\pageref{firstpage}--\pageref{lastpage}} \pubyear{2010}

\maketitle

\label{firstpage}

\begin{abstract}
  An ionization front expanding into a neutral medium can be slowed-down significantly by recombinations. In cosmological numerical simulations the recombination rate is often computed using a \lq clumping factor\rq\, that takes into account that not all scales in the simulated density field are resolved. Here we demonstrate that using a single value of the clumping factor significantly overestimates the recombination rate, and how a local estimate of the clumping factor is both easy to compute, and gives significantly better numerical convergence. We argue that this lower value of the recombination rate is more relevant during the reionization process and hence that the importance of recombinations during reionization has been overestimated.
\end{abstract}

\begin{keywords}
radiative transfer -- methods:numerical -- cosmology: dark ages, reionization, first stars
\end{keywords}

\section{Introduction}
\label{intro}
The intergalactic medium is highly ionized by the UV-background
produced by QSOs and galaxies, at least since $z\sim 6$
\citep[][]{gunn65, haardt96, rauch98}. Reionization -- the transition
from neutral to ionized -- occurred around $z_\mr{reion} \sim 10$,
according to the Thomson optical depth inferred from the
cosmic-microwave background \citep[see][for 7-year WMAP
result]{komatsu10}.  Reionization starts when the first sources of
ionizing photons form small, isolated H{\sc II} regions around them. As more and increasingly luminous sources form, ionized regions become
larger and more numerous, until they eventually percolate space,
signalling the end of the epoch of reionization
\citep[EoR,][see recent reviews by \textit{e.g.} \citealt{barkana01,
  ciardi05, loeb06}]{arons72,haardt96,giroux96,gnedin97,haiman97}.

The nature of the sources of ionizing photons is currently unknown,
with \lq first stars\rq, early galaxies, and black hole accretion,
probably all contributing to some extent \citep[\textit{e.g.}][]{madau99}. We recently demonstrated that the early population of galaxies predicted by Durham's \galform\ model produce enough ionizing photons to complete reionization by $z\sim 10$ \citep{Raicevic10a}. The same model also matches very well the
luminosity function of redshift $z=7-10$ galaxies recently discovered
by the {\em Hubble Space Telescope} \citep{Bouwens08, Bouwens09}. In
this model, the bulk of photons are produced in low-mass ($M_\star\sim
10^6\,h^{-1}M_\odot$), gas-rich, faint galaxies (rest-frame UV
magnitude $M_{1500, {\rm AB}}\sim -16$) during a star bursts ($\dot
M_\star\sim 0.04\,h^{-1}M_\odot$~yr$^{-1}$) induced by a merger.

The fraction $\cal R$ of photons emitted per HI necessary to complete
reionization is ${\cal R} = (1 + N_\mr{rec}) / f_\mr{esc}$, where
$f_\mr{esc}$ is the fraction of ionizing photons that can escape their
host galaxies and $N_\mr{rec}$ is the average number of recombinations
per HI. We distinguish between recombinations that occur in ({\em i})
mini-haloes, ({\em ii}) Lyman-limit systems (LLS), and ({\em iii}) in
the general intergalactic medium (IGM).  Mini-haloes are small
high-density clouds that are too cold to form stars via atomic line
cooling. The presence of mini-halos can have a significant effect on
the propagation and size of HII regions during reionization
\citep{Furlanetto05, mcquinn07}. However, when overrun by an ionization front their gas will be photo-heated and they will eventually evaporate \citep{shapiro04,iliev05-1,ciardi06}, which decreases their importance in the later stages and after reionization. On the other hand, Lyman-limit systems have sufficiently high column-densities, $N_{\rm HI} \gtrsim 10^{17}$~cm$^{-2}$, to self-shield. Ionizing photons impinging on such optically thick systems get converted to (non-ionizing) Lyman-$\alpha$ radiation at high-efficiency \citep{hogan1987,rauch1998}. These larger systems determine the mean free path of ionizing photons in the post-reionization era \citep[\textit{e.g.}][]{miralda-escude03}. Below we will concentrate on the third source of recombinations, those occurring in the IGM.

The recombination rate per unit volume, $\dot n_\mr{rec}$, depends on the particle density squared, $\dot n_\mr{rec}=\alpha x^2\,n^2$, where $x$ is the mean ionized fraction, and hence varies thorough out the inhomogeneous IGM. Early semi-analytical models of reionization used a mean recombination rate, with the IGM inhomogeneity expressed in terms of a \lq clumping factor\rq\ ${\cal C}$, $\langle \dot n_\mr{rec}\rangle=\alpha x^2\,\langle n^2\rangle\equiv \alpha\,x^2{\cal C}\,\langle n\rangle^2$ \citep[\textit{e.g.}][]{giroux96, tegmark97, ciardi97, haiman97, madau99, valageas99}. Numerical simulations of \cite{gnedin97} yielded high estimates of ${\cal C}\sim 10$ (40) at redshifts $z=8$ (5), implying that recombinations are generally quite important. Similar values have been used in the estimate of the comoving star formation rate density needed to keep the post-reionization Universe ionized by balancing ionizations with recombinations. The inferred value,
\begin{eqnarray}
\dot{\rho}_\star &\approx& 0.03 \, M_{\odot} \, {\rm yr}^{-1}\,{\rm  Mpc}^{-3} \, \nonumber \\ 
&\times& {1\over f_{\rm esc}} \, \left({1+z \over 8} \right)^3 \,  \left({\Omega_b\,h_{70}^2 \over 0.0457}\right)^2 \, \left({{\cal C} \over 30}\right)\,,
\label{eq:sfr}
\end{eqnarray}
\citep[\textit{e.g.} ][]{madau99} is significantly higher at $z \sim 7$ than some currently observationally inferred rates \citep[\textit{e.g.}][]{bunker10}.

Current numerical simulations model reionization by directly following
the propagation of ionization fronts through the inhomogeneous IGM,
thus in principle eliminating the need for a clumping factor
\citep[\textit{e.g.}][]{sokasian01, ciardi03, iliev06, mcquinn07,trac07}. However the resolution of the radiative transfer (RT) calculation is in general much coarser than that of the density field on which the sources are identified \citep[e.g.][]{Iliev10}, and a single clumping factor, usually derived from higher resolution small box runs, is used to take account of the density structure below the resolution of the RT mesh \citep[\textit{e.g.}][]{ciardi03, iliev07}.

In this letter we will show that a single value of the clumping factor
is, in fact, not appropriate for estimating the recombination rate in
numerical RT reionization simulations, and leads to a significant over
estimate of the importance of recombinations. 

\section{Definition of the clumping factor}
\label{method}
The recombination rate of Hydrogen-only gas in a volume $V$ is
\begin{eqnarray}
\dot{N}_{\rm rec} &=& \alpha \, \int_V \, \nh^2 \dd V \nonumber \\
                 &\equiv& \alpha \, {\cal C}\, \langle \nh \rangle^2 \, V\,,
\label{eq:rec}
\end{eqnarray}
where $\alpha$ is the recombination coefficient, $\nh$ the hydrogen number density, ${\cal C} \equiv \langle \nh^2\rangle/\langle \nh \rangle^2$ the clumping factor, and we
assumed the gas to be fully ionized $(\nh = \nhii)$; the angular brackets
denote a volume average.

Most current numerical models of reionization follow the formation of
dark matter structures in a cosmological setting, compute emissivities
of galaxies associated with dark matter halos, then follow how these
galaxies ionize their surroundings with a radiative transfer
calculation \citep{ciardi03,iliev06,mcquinn07,trac07, Raicevic10b}. Our method in this letter is designed to reproduce the steps taken to set up RT computational meshes in such simulations.

We use a dark matter simulation performed with the Tree-PM code Gadget-2 \citep{springel05}. The simulation uses $1024^3$ equal-mass particles in a periodic cosmological volume of size 20~$h^{-1}$ comoving Mpc, assuming a flat $\Lambda$CDM cosmology with cosmological parameters [$\Omega_m, \, \Omega_b, \, \Omega_\Lambda, \,
h, \, \sigma_8, \, n_s]$~=~[0.25, 0.045, 0.75, 0.73, 0.9, 1]. Baryons
are assumed to trace the dark matter, and hence the gas density
$\rho_g$ is related to the matter density $\rho$ as $\rho_{\rm
  g}=(\Omega_b/\Omega_m)\,\rho$. The matter density $\rho$ at the
position ${\bf r}_i$ of particle $i$ is estimated using the SPH
algorithm \citep{lucy79,gingold79},
\begin{equation}
\rho({\bf   r}_i)=\sum_j\,m_{j} W({|{\bf r}_i-{\bf r}_j|\over
  h_i})\,,\label{eq:SPH}
\end{equation}
where $m_i$ is the particle mass, and $h_i$ its \lq resolution
length\rq\, chosen so that it holds $\sim 40$ neighboring particles $j$ that contribute to the sum; $W$ is the smoothing kernel. The (Hydrogen)
number density is computed as $\nh=(1-Y)\,\rho_g/m_{\rm p}$, where $Y$ is
the primordial Helium abundance by mass, and $m_{\rm p}$ is the proton
mass.

Assigning a volume $V_i \approx m_i/ \rho_i$ to particle $i$, allows us
to compute the mean clumping factor as
\begin{equation}
{\cal C}=\frac{1}{N_\mr{part}^2} \, \sum_i^{N_\mr{part}} \,n_{\mr{H},i} \, \sum_i^{N_\mr{part}}{1\over n_{\mr{H},i}}\, , \mbox{ for } n_{\mr{H},i} \leq n_\mr{thr}\,,
\label{eq:c}
\end{equation}
where $N_\mr{part}$ is the number of particles in volume $V$ with
number density lower than a given threshold density $n_\mr{thr}$. We use
$n_\mr{thr} \equiv \Delta_\mr{thr} \langle \nh \rangle$ with $\Delta_{\rm
  thr}=100$, to exclude
collapsed halos from the IGM density field
\citep[see \textit{e.g.}][]{miralda-escude00, miralda-escude03,pawlik09}. Baryons in
halos do not trace the dark matter but collapse to form galaxies. Due
to their high densities, galaxies should not be treated as general IGM,
but rather as LLS and their effect on the propagation of ionizing
radiation described in terms of a mean free path, as in \textit{e.g.} \citet{madau97}. We chose the overdensity threshold of
$\Delta_\mr{thr} = 100$ appropriate for the density at the virial
radius of a halo, and also to allow for a direct comparison to other
works \citep[\textit{e.g.}][]{pawlik09}.

Finally, the RT calculations for large-scale reionization models are
usually performed on a uniform cubic mesh, with the density at each
mesh point obtained from the $N$-body particles using, for example,
nearest grid point interpolation \citep[\textit{e.g.}][]{hockney88}. We employ
several such grids in the following discussion. The details of the
particular simulation we use, and the way we compute densities,
are not important as far as our conclusions on recombinations are
concerned.

\section{The local clumping}
\begin{figure}
  \begin{center}
    \includegraphics[width=0.48\textwidth,clip=true, trim=10 10 30 30, keepaspectratio=true]{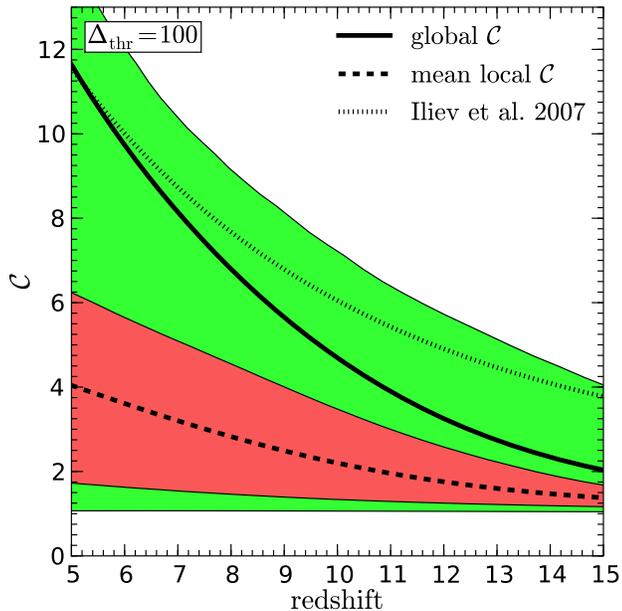}
  \end{center}
  \caption{Evolution of the clumping factor in a simulation with
    1024$^3$ particles in a 20~$h^{-1}$~Mpc box, neglecting particles
    with overdensities higher than a threshold value of $\Delta_{\rm
      thr}=100$. The {\em black solid line} shows the global clumping
    factor, which follows reasonably well the result obtained by Iliev
    et al. (2007; {\em black dotted line}. We then divide the volume in
    64$^3$ equal sub-volumes, and compute the local clumping factor in
    each of them. The {\em dashed line} shows the mean local clumping
    factor, with the $50\%$ ($99\%$) percentiles indicated by the red
    (green) shaded region.  Clearly there is a large scatter in ${\cal
      C}_{\rm local}$, and its mean value is significantly lower than
    that of the global clumping factor.}
  \label{fig1}
\end{figure}

\begin{figure}
  \begin{center}
    \includegraphics[width=0.45\textwidth, clip=true, trim=25 20 40       40, keepaspectratio=true]{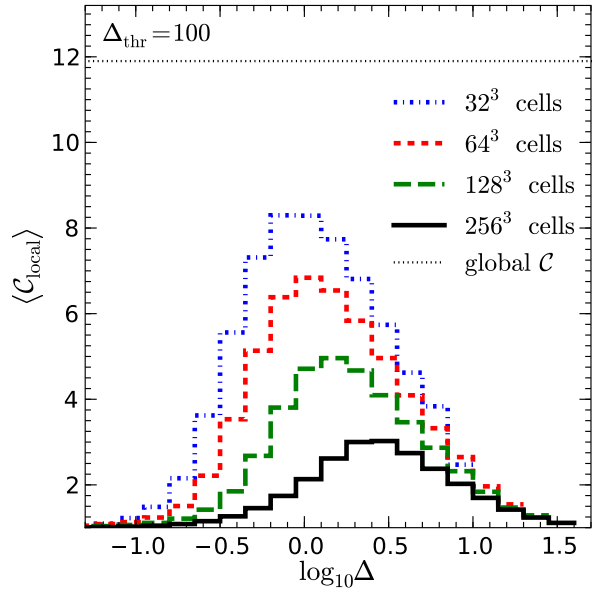}
  \end{center}
  \caption{Mean local clumping factor ($\langle{\cal C}_{\rm local}\rangle$) as a
    function of the overdensity ($\Delta$) of the sub-volume, at redshift $z=5$,
    for various subdivisions of the computational volume; the
    corresponding cell sizes are 625, 312.5, 156.25 and
    78.125~$h^{-1}$~ kpc. Coarser sub-cells yield higher values of 
    $\langle{\cal C}_{\rm local}\rangle$ at a given overdensity, and
    $\langle{\cal C}_{\rm local}\rangle$ peaks at intermediate values
    of the overdensity. The value of $C_\mr{global}$ is shown as a
    black dotted line.}
  \label{fig2}
\end{figure}

We will make a distinction between two types of clumping factors in the following discussion, both based on Eq.~(\ref{eq:c}). The \textit{global} clumping factor, ${\cal C}_\mr{global}$, is computed by summing over {\em all} particles, as is done in \textit{e.g.} \citet{iliev05, iliev07} and \citet{pawlik09}. However we can also divide the computational volume in (equal volume, non-overlapping) sub-volumes, {\em i.e.} a uniform mesh\footnote{The choice of the volume subdivision is motivated by simplicity. Our conclusions are not dependent on the shape of the mesh}, and evaluate ${\cal C}$ in each sub-volume (mesh cell) by summing only over particles in that sub-volume\footnote{Note that in general the mean density in each sub-volume will be different as well.}. We will call this a \textit{local} clumping factor, ${\cal C}_\mr{local}$. 

We compare ${\cal C}_\mr{global}$ and ${\cal C}_\mr{local}$ computed on $64^3$ equal sub-volumes of our 1024$^3$ particles, 20~$h^{-1}$Mpc aside simulation  box in Fig.~\ref{fig1}. Our value for ${\cal C}_{\rm global}$ is in reasonable agreement with that obtained by \cite{iliev07}, even though the relations are derived from significantly different N-body runs. Interestingly however, at any $z$, there is a large range of values of ${\cal C}_{\rm local}$, up to more than a factor of 10 at $z=5$. In addition also the mean value of ${\cal C}_{\rm local}$ in all sub-volumes of the simulation box is significantly lower than that of ${\cal C}_{\rm global}$. Clearly, a single value of ${\cal C}$ is not able to characterize the recombination rate in every region of a simulation.

The mean ${\cal C}_{\rm local}$ in sub-volumes with a given overdensity $\Delta$ is shown in Fig. \ref{fig2}. This type of ${\cal C}(\Delta)$ relation was not previously discussed in the literature. The highest values of ${\cal C}_{\rm local}$ are found in sub-volumes with intermediate overdensity, because clumping is a measure of \textit{inhomogeneity} in the density field. Therefore high clumping is found in sub-volumes that contain high gradients in the density field, for example, small high-density halos in a low density region, or the transition between a filament and a void. Improving the sampling of the density field by using smaller sub-volumes, decreases the peak amplitude in ${\cal C}_{\rm local}$, and moves the maximum toward higher overdensity regions, but does not change the general shape of the distribution. The general shape is also not strongly affected by the choice of $\Delta_\mr{thr}$. Note that the recombination rate, which is $\propto {\cal C}_{\rm local}\Delta^2$, is still highest in the highest density regions. However, the use of ${\cal C}_\mr{local}$ distribution shown in Fig. \ref{fig2} increases the relative contribution of moderately overdense regions to the total recombination rate.

\begin{figure}
  \begin{center}
    \includegraphics[width=0.45\textwidth,keepaspectratio=true,clip=true, trim=20 15 30 25]{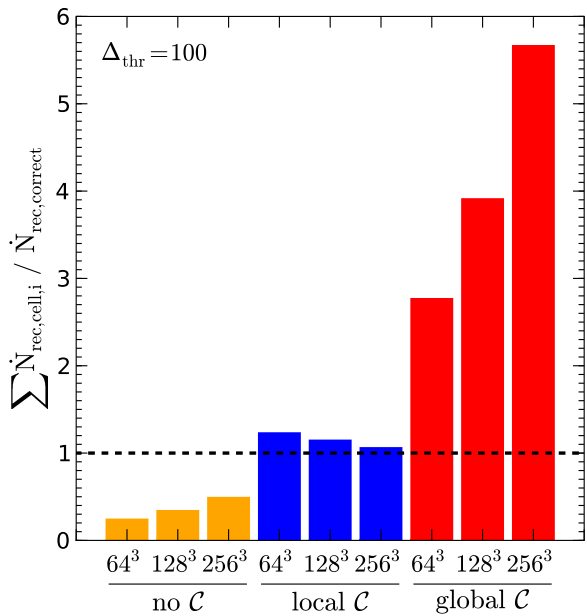}
  \end{center}
  \caption{Recombination rate, $\dot{N}_\mr{rec}$, in a fully-ionized
    simulation box of 20~$h^{-1}$Mpc at redshift $z=5$ when simulated
    using 1024$^3$ particles, assuming gas traces dark matter, and
    imposing an overdensity threshold of $\Delta_{\rm thr}=100$.
    The recombination rate is computed using various sizes of the
    sub-volumes over which to compute the clumping factor, and is
    expressed in units of the \lq correct\rq\ value computed directly
    from the $N$-body particles. Not using a clumping factor at all
    (${\cal C}=1$; yellow bars) leads to an underestimate of $\dot
    N_{\rm rec}$, which get less at improved sub-sampling of the
    density field. Using a global value (${\cal C}_{\rm global}$; red
    bars) leads to a large overestimate of $\dot N_{\rm rec}$, which
    gets {\em worse} at improved sampling. Finally using a locally    estimated clumping factor (${\cal C}_{\rm local}$; blue bars) gives
    a much more accurate value of $\dot N _{\rm rec}$, which is also
    nearly independent of the sampling resolution.}
  \label{fig3}
\end{figure}

An inaccurate estimate of the clumping factor of course also implies
that the recombination rate is not accurate, and hence that the speed
of ionisation fronts are not computed correctly.  Instead of studying the effects on a full RT reionization run, we show a simpler example by calculating recombination rates in the 20$h^{-1}$~Mpc, 1024$^3$ particles simulation box at redshift $z=5$, including densities up to a threshold of $\Delta_{\rm thr}=100$ and assuming that all gas is fully ionised.  We can sum the recombination rate per particle, $\dot N_{{\rm rec},i}=\alpha\,(m/m_{\rm p}^2)\,\rho_{\mr{g},i}$, over all particles $i$, to get the total recombination rate, $\dot N_\mr{rec,correct} = \sum \dot{N}_{{\rm rec},i}$. This is the \lq correct\rq\
recombination rate as it takes into account all the available density field information from the N-body simulation.  We compare the value of $\dot N_\mr{rec,correct}$ to values obtained by first interpolating particle densities to a mesh as described in Section 2, and computing the sum of  $\dot N_{\rm rec}$ in each mesh cell using different expressions for the clumping factor (Fig. \ref{fig3}). Not using a clumping factor (yellow bars) shows the effect of the density field smoothing by the mesh as the recombination rates are both underestimated (a factor of two even at the highest grid resolution) and strongly mesh resolution dependent. On the other hand, when the global clumping ${\cal C}_\mr{global}$ is used to represent the sub-grid matter distribution (red bars), it significantly overestimates the recombination rate while not remedying the dependence on mesh resolution. This is no surprise as a single value of clumping simply linearly increases the no clumping result. Crucially, using the locally estimated clumping factor, ${\cal C}_{\rm local}$ (blue bars), leads to a much more accurate and resolution independent description of recombinations (within $\sim 25\%$ of $\dot{N}_\mr{rec,correct}$ on all grid resolutions). Note that the assumption of a fully ionized simulation box is a special case, chosen for illustrative purposes. A more relevant case for the study of reionization is a partially ionized cosmological density field. The most important consequence of using ${\cal C}_\mr{local}$ instead of ${\cal C}_\mr{global}$ is a much lower average recombination rate, for whichever overdensity regions are ionized at any time.  The convergence of recombination rates obtained in Fig. \ref{fig3} with the use of ${\cal C}_\mr{local}$ also leads to the convergence of I-front speeds during reionization as we will show in \cite{Raicevic10b}, where we also discuss the ionized fraction as a function of overdensity during reionization. Also, ${\cal C}_\mr{local}$ is computed assuming a full ionization and does not provide a perfectly accurate recombination rate estimate for partially ionized sub-volumes. Therefore, the use of a higher resolution RT computational mesh is always preferable to employing any kind of clumping.

\section{Conclusions}
\label{discussion}

Clumping factors are often used in simulations of reionization to
represent the unresolved matter inhomogeneity, below the computational mesh resolution, which may significantly contribute to the recombination rates \citep[\textit{e.g.}][]{ciardi03, iliev07, mcquinn07}. Here we have shown that, because the density field during reionization is so inhomogeneous, using a single clumping factor in general leads to a significant over estimate of the recombination rate (factors of several), which may in fact get worse as the grid resolution is improved. As a consequence the speed of ionization fronts is artificially depressed, reionization delayed, and the outcome of the radiative transfer simulations significantly resolution
dependent.

\cite{mcquinn07} improved on this issue by analytically deriving the clumping factor as a function of overdensity, ${\cal
  C}={\cal C}(\Delta)$, and also \cite{kohler07} took into account some density dependence by splitting their simulation volumes into 8 equal sub-volumes. However, both ignored the fact that clumping still depends on the volume over which it is computed ({\em i.e.} the size of the sub-volumes), and that dependence is not negligible as shown in Fig. \ref{fig2}. Not taking the volume into account will always lead to the RT results depending on the computational mesh resolution.

Fortunately it is not computationally intensive to compute a local
clumping factor on a reasonably fine grid. The recombination rate
obtained using this clumping factor is very close to the \lq
correct\rq\ value inferred directly form the particles themselves, and
is not very sensitive to the resolution of the mesh used
(see Fig.~\ref{fig3}). In our dark matter only simulations the cell size should be such that most of the structures are resolved, and should ideally be close to the Jeans mass of the photo-ionised IGM
\citep{pawlik09}. We therefore argue that any future fits of clumping
factors aimed for use in RT simulations of reionization must include
both density and volume dependence. The alternative of computing
the recombination rate per particle directly \citep{trac07}, is
far more computationally expensive.

\section*{Acknowledgments}

MR would like to thank Garrelt Mellema and Joop Schaye for useful discussions. During the work on this paper, MR was
supported by a grant from Microsoft Research Cambridge.

\bibliographystyle{mn2e}
\bibliography{main}

\label{lastpage}

\end{document}